\documentclass[12pt]{iopart}

\expandafter\let\csname equation*\endcsname\relax
\expandafter\let\csname endequation*\endcsname\relax

\usepackage{epsfig}
\usepackage{amsmath}
\usepackage{amsfonts}
\usepackage{amssymb}
\usepackage{graphicx}

\begin{document}

\begin{flushright}
 {\tt BROWN-HET-1640}
\end{flushright}

\title{Perturbative and Non-Perturbative Aspects in Vector Model / Higher Spin Duality}

\author{Antal Jevicki$^1$, Kewang Jin$^2$ and Qibin Ye$^1$}

\address{$^1$ Department of Physics, Brown University, Providence, RI 02912, USA \\
$^2$ Institut f\"{u}r Theoretische Physik, ETH Z\"{u}rich, CH-8093 Z\"{u}rich, Switzerland}

\eads{\mailto{antal\_jevicki@brown.edu}, \mailto{jinke@itp.phys.ethz.ch}, \mailto{qibin\_ye@brown.edu}}

\begin{abstract}

We review some recent work on AdS/CFT duality involving the 3d $O(N)$ Vector Model and AdS$_4$ Higher Spin Gravity. Our construction is based on bi-local collective field theory which provides an off-shell formulation of Higher Spin Gravity with $G=1/N$ playing the role of a coupling constant. Consequently perturbative and non-perturbative issues of the theory can be studied. For the correspondence based on free CFT's we discuss the nature of bulk $1/N$ interactions through an $S$-matrix which is argued to be equal to 1 (Coleman-Mandula theorem). As a consequence in this class of theories nonlinearities are removable, through a nonlinear field transformation which we show at the exact level. We also describe a geometric (K\"{a}hler space) framework for the bi-local theory which applies equally simple to $Sp(2N)$ fermions and the de Sitter correspondence. We discuss in this framework the structure and size of the bi-local Hilbert space and the implementation of (finite $N$) exclusion principle.

\end{abstract}

\submitto{\JPA \\ Special volume on ``Higher Spin Theories and AdS/CFT'' \\
Edited by M. R. Gaberdiel and M. Vasiliev}
\maketitle

\section{Introduction}

The duality involving large $N$ vector models and Vasiliev's Higher Spin Gravity in Anti de Sitter space represents one of the simplest examples of AdS/CFT correspondence and has been a topic of vigorous recent studies \cite{Klebanov:2002ja, Sezgin:2002rt, Petkou:2003zz, Das:2003vw, Giombi:2009wh, Giombi:2010vg, Koch:2010cy, Douglas:2010rc, Jevicki:2011ss, Shenker:2011zf, Aharony:2011jz, Giombi:2011kc, Maldacena:2011jn, Bekaert:2012ux, Vasiliev:2012vf, Maldacena:2012sf, deMelloKoch:2012vc}. Equally interesting is a lower dimensional correspondence between 2d minimal CFT's and 3d Higher Spin Gravities \cite{Henneaux:2010xg, Campoleoni:2010zq, Gaberdiel:2010pz, Gaberdiel:2011zw, Chang:2011mz, Gutperle:2011kf, Ammon:2011nk, Kraus:2011ds, Castro:2011fm, Ammon:2011ua, Gaberdiel:2012yb, Gaberdiel:2012ku}. These large $N$ dualities involve some fairly well understood quantum field theories  and a relatively novel version of Gravity built on a single Regge trajectory. These theories \cite{Vasiliev:1995dn, Vasiliev:2003ev, Bekaert:2005vh, Metsaev:2005ar, Metsaev:2007rn, Polyakov:2010sk, Iazeolla:2011cb} feature many properties that have been unreachable in String Theory, in particular the structure and explicit form of the higher spin gauge group. They also offer a potentially solvable framework for study of black holes \cite{Gutperle:2011kf, Ammon:2011nk, Kraus:2011ds, Castro:2011fm, Gaberdiel:2012yb} and de Sitter theory \cite{Anninos:2011ui, Ng:2012xp, Das:2012dt}.

In the case of three dimensional $O(N)$ vector field theory, one has two conformally invariant fixed points, the UV and the IR one. The HS duals are given by the same Vasiliev theory but with different boundary conditions on the scalar field \cite{Klebanov:2002ja}. This provides a simple relationship between a (HS) theory dual to a free $N$-component scalar field (UV) and the nontrivial dual corresponding to the IR CFT. The correspondence provided by the free $O(N)$ scalar field theory is then of central interest. This theory is characterized by an infinite sequence of conserved currents which are themselves boundary duals of higher spin fields and whose correlation functions represent a point of comparison \cite{Giombi:2009wh, Giombi:2010vg} between the two descriptions.

The collective field approach \cite{Das:2003vw, Koch:2010cy} that we will review provides a construction of bulk HS Gravity and AdS space-time in terms of composite  bi-local fields. It leads to an effective nonlinear bulk theory (with $1/N$ appearing as the coupling constant) which was seen to possess all the properties of the dual AdS theory. This theory reproduces arbitrary-point correlation functions of the HS theory (in various gauges) \cite{Jevicki:2011ss}. It provides an explicit  mechanism for the emergence of AdS space-time and of Higher Spin Gravity degrees of freedom through collective effects. These effects and the duality with Vasiliev's theory is there already for free CFT's since one has a nontrivial large $N$ system. The question then arises regarding the implementation of the Coleman-Mandula theorem \cite{Maldacena:2011jn, Maldacena:2012sf} with respect to the triviality of bulk interactions governed by the $G=1/N$ parameter. As we explain the bi-local construction offers a framework for defining and calculating an $S$-matrix and addressing in this way the implementation of the Coleman-Mandula theorem. This was done in \cite{deMelloKoch:2012vc} with the result $S=1$ implying triviality of bulk interactions, which can be removed by a nonlinear field transformation. A construction of the linearizing transformation was given in \cite{deMelloKoch:2012vc} and it will be briefly reviewed.

In the second part of this short overview we describe a geometric, pseudo-spin representation \cite{Das:2012dt} of the theory. Perturbatively this representation is entirely equivalent to the bi-local collective field one, however at the non-perturbative level it leads to further insights. This concerns the question of the quantization of the coupling constant and the implementation of the finite $N$ exclusion principle and cutoff. The formulation in terms of pseudo-spins applies equally well to fermionic and bosonic large $N$ systems. It in fact exhibits an explicit relationship between the two which  corresponds to a change of a sign \cite{Das:2012dt} in the coupling constant $G$, or simply of $N$ into $-N$. The pseudo-spin representation is also seen to have a geometric meaning, namely corresponding to a dynamics on matrix-valued Grassmanian (K\"{a}hler) manifolds. Consequently methods of geometric quantization can be employed in defining the Hilbert space of the theory. This leads to a very satisfactory result regarding the number of states in the Hilbert space. When dealing with bi-local composite variables one point of concern is the fact that this set (at finite $N$) appears to be over-complete. Consequently there is the legitimate worry that the number of states in the corresponding (bi-local) Hilbert space will itself be over-complete. However the geometric (K\"{a}hler) nature of the bi-local space together with geometric quantization is seen to faithfully represent the original $N$-component theory. This issue is of definite importance for Black Hole Entropy and reconstruction of de Sitter space itself.

\section{From $O(N)$ to Bi-local Field Theory }

In constructing $O(N)$/Higher Spin duality one starts with a $N$-component scalar field theory
\begin{eqnarray}
\mathcal{L}= \frac{1}{2}\partial_\mu \phi^a \partial^\mu \phi^a+\frac{g}{4}(\phi\cdot\phi)^2 \ , \qquad a=1,\cdots,N
\end{eqnarray}
where $\phi^a=\phi^a(t,\vec{x})=\phi^a(x^+,x^-,x^\perp)$. This theory features two critical points with conformal symmetry: the UV fixed point at zero coupling ($g=0$) and the nontrivial IR fixed point at nonzero value of the coupling constant. The latter represents the celebrated case of three dimensional critical phenomena. Both cases have a nontrivial large $N$ limit which is solvable.

In the correspondence with higher spin fields, at the free UV fixed point, a central role is played by the sequence of traceless and symmetric higher spin currents
\begin{equation}
J_{\mu_1 \mu_2 \cdots \mu_s}=\sum_{k=0}^s (-1)^k 
\begin{pmatrix} s-1/2 \cr k \end{pmatrix}
\begin{pmatrix} s-1/2 \cr s-k \end{pmatrix}
\partial_{\mu_1} \cdots \partial_{\mu_k} \phi^a \;
\partial_{\mu_{k+1}} \cdots \partial_{\mu_s} \phi^a -\text{traces}
\end{equation}
which are exactly conserved. These operators can be summed up by the semi bi-local generating functional
\begin{eqnarray}
\mathcal{O}(x,\epsilon)=\phi^a(x-\epsilon)\sum_{n=0}^\infty \frac{1}{(2n)!}
\Bigl(2\epsilon^2 \overleftarrow{\partial_x} \cdot \overrightarrow{\partial_x}
-4(\epsilon \cdot \overleftarrow{\partial_x})(\epsilon \cdot \overrightarrow{\partial_x})\Bigr)^n
\phi^a(x+\epsilon)
\end{eqnarray}
where $\epsilon^2=0$ as to satisfy the traceless condition. Consequently, $\epsilon$ represents a two dimensional space (a cone) and altogether $\mathcal{O}(x,\epsilon)$ is a five dimensional semi bi-local field. The currents that it generates  are (boundary) duals of AdS$_4$ higher spin fields
\begin{eqnarray}
J_{\mu_1 \mu_2 \cdots \mu_s}(x) \longleftrightarrow \mathcal{H}_{\hat{\mu}_1 \hat{\mu}_2 \cdots \hat{\mu}_s}(x,z \rightarrow 0)
\end{eqnarray}
where in Poincar\'{e} coordinates $ds^2=\tfrac{1}{z^2}(dx^2+dz^2)$ the boundary is at $z=0$. In the AdS/CFT correspondence, correlation functions of currents are to match up with the boundary transition amplitudes. This collection of correlation functions is sometimes referred as the boundary $S$-matrix of the AdS theory. A successful comparison between the two was accomplished in the three-point case by Giombi and Yin \cite{Giombi:2009wh, Giombi:2010vg} who were able to match the two critical points of the vector model with two boundary versions of Vasiliev's Higher Spin Gravity in AdS$_4$. The trivial and nontrivial fixed points are seen as conjectured by Klebanov and Polyakov \cite{Klebanov:2002ja} to correspond to different boundary conditions involving the lowest spin ($s=0$) scalar field.

The constructive approach for this AdS$_4$/CFT$_3$ duality in terms of collective fields was formulated in \cite{Das:2003vw}. It is based on all $O(N)$ invariant bi-local fields
\begin{eqnarray}
\Phi(x,y)\equiv \phi(x)\cdot\phi(y)=\sum_{a=1}^N\phi^a(x)\cdot\phi^a(y)
\end{eqnarray}
which have the property that they close under the Schwinger-Dyson equations (in the large $N$ limit). These operators represent a more general set than the conformal currents $\mathcal{O}(x,\epsilon)$ since there is no restriction to the cone. They encompass one extra dimension and are consequently natural candidates for representing the bulk AdS$_4$ theory. This proposal is supported by the fact that there is an effective, collective field action with the property that it exactly evaluates the $O(N)$ singlet partition function
\begin{eqnarray}
Z=\int [d\phi^a(x)] e^{-S[\phi]}=\int \prod_{x,y} [d\Phi(x,y)] \mu(\Phi) e^{-S_c[\Phi]}
\end{eqnarray}
where the measure is given by $\mu(\Phi)=(\det \Phi)^{V_x V_p}$ with $V_x=L^3$ the volume of space and $V_p=\Lambda^3$ the volume of momentum space where $\Lambda$ is the momentum cutoff. Explicitly the (collective) action reads
\begin{eqnarray}
S_c[\Phi]=\text{Tr}\left[-(\partial_x^2+\partial_y^2)\Phi(x,y)+V \right]+\frac{N}{2}\text{Tr} \, \ln \Phi(x,y)\label{action}
\end{eqnarray}
where the trace is defined as ${\rm Tr} \, B=\int d^3 x \, B(x,x)$. Here one has a natural (star) product defined as $(\Psi \star \Phi)(x,y)=\int dz \, \Psi(x,z)\Phi(z,y)$. 

The collective action features $1/N$ as the expansion parameter even when the original $N$-component theory one considers is free. Note the CFT interactions only appear in the collective representation as linear or quadratic terms. Through the identification of $1/N$ with $G$ (the coupling constant of higher spin gravity), this collective field representation simply provides a bulk description of the dual Higher Spin theory in AdS space-time. 

Calculations and the perturbative ($1/N$) expansion in this (bi-local) theory proceeds in the standard way. From the equations of motion specified by $S_c$ one obtains the background $\Phi_0$ and the shift: $\Phi=\Phi_0+\frac{1}{\sqrt{N}}\eta$ induces a sequence of interaction vertices \cite{deMelloKoch:1996mj} 
\begin{eqnarray}
S_c[\Phi]=S_c[\Phi_0]+\text{Tr}[\Phi_0^{-1} \eta \Phi_0^{-1} \eta]+\frac{g}{4}\eta^2+\sum_{n \ge 3} N^{1-n/2} \, \text{Tr} \, B^n 
\end{eqnarray}
where $B \equiv \Phi_0^{-1} \eta$. The nonlinearities built into $S_c$ are precisely such that all invariant correlators: $\langle \phi(x_1) \cdot \phi(y_1) \cdots \phi(x_n) \cdot \phi(y_n)\rangle$ are now reproduced through the Witten (Feynman) diagrams with $1/N$ vertices. Again we stress that this nonlinear structure is there for both the free UV and the interacting IR fixed points.

\begin{figure}
\begin{center}
\includegraphics[width=0.6\textwidth]{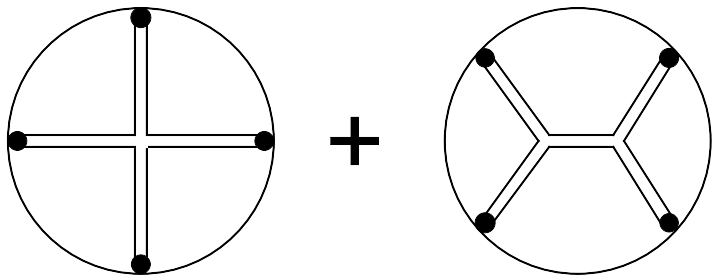}  
\caption{Illustration of the four-point collective field diagrams.}
\label{fourdiagram}
\end{center}
\end{figure}

This bi-local theory is expected to represent a (covariant) gauge fixing of Vasiliev's gauge invariant theory. A large number of degrees of freedom are removed in fixing a gauge in any gauge theory and this simplification happens in Higher Spin Gravity too. A one-to-one relationship between bi-local and AdS higher spin fields can be demonstrated in a physical (single-time) picture. The existence of such a gauge and the discussion of the collective dipole underlying the collective construction was given in \cite{Jevicki:2011ss}.

The single-time formulation involves the equal time bi-locals
\begin{eqnarray}
\Psi(t,\vec{x},\vec{y})=\sum_{a=1}^N \phi^a(t,\vec{x})\phi^a(t,\vec{y}) \label{bilocal}
\end{eqnarray}
and their conjugates: $\Pi(\vec{x},\vec{y})=-i\frac{\delta}{\delta\Psi(\vec{x},\vec{y})}$ with the effective Hamiltonian given as
\begin{eqnarray}
H=2\text{Tr}(\Pi \Psi \Pi)+\frac{1}{2}\int d\vec{x} [-\nabla_{\vec{x}}^2\Psi(\tilde{x},\tilde{y})\vert_{\tilde{x}=\tilde{y}}]+\frac{N^2}{8}\text{Tr} \, \Psi^{-1} \ . \label{H}
\end{eqnarray}
Again, as in the case of the action, nontrivial CFT interactions are simply added as composite operators. This Hamiltonian again has a natural $1/N$ expansion, after a background shift
\begin{eqnarray}
\Psi=\Psi_0+\frac{1}{\sqrt{N}} \eta \ , \qquad \Pi=\sqrt{N} \pi \ ,
\end{eqnarray}
the quadratic Hamiltonian reads
\begin{eqnarray}
H^{(2)}=2 {\rm Tr}( \pi \Psi_0 \pi )+\frac{1}{8} {\rm Tr} (\Psi_0^{-1} \eta \Psi_0^{-1} \eta \Psi_0^{-1}) \ .
\end{eqnarray}
Fourier transforming the background field as well as the fluctuations (see \cite{deMelloKoch:2012vc} for more details), one finds in momentum space
\begin{equation}
H^{(2)} = \frac{1}{2} \int d \vec{k}_1 d \vec{k}_2 \; \pi_{\vec{k}_1 \vec{k}_2} \pi_{\vec{k}_1 \vec{k}_2}
+ \frac{1}{8} \int d \vec{k}_1 d \vec{k}_2 \; \eta_{\vec{k}_1 \vec{k}_2}
\left( \psi_{\vec{k}_1}^{0\,\,-1}+\psi_{\vec{k}_2}^{0\,\,-1} \right)^2\eta_{\vec{k}_1 \vec{k}_2}
\end{equation}
which correctly describes the (singlet) spectrum of the $O(N)$ theory with $\omega_{\vec{k}_1 \vec{k}_2} = \sqrt{\vec{k}_1^2} + \sqrt{\vec{k}_2^2}$. Higher vertices representing interactions $1/N$ can be found similarly, in particular, the cubic and quartic interactions are given explicitly as
\begin{eqnarray}
H^{(3)}&=&\frac{2}{\sqrt{N}}\text{Tr}(\pi \eta \pi)
-\frac{1}{8\sqrt{N}}\text{Tr} (\Psi_0^{-1} \eta \Psi_0^{-1} \eta \Psi_0^{-1} \eta \Psi_0^{-1}) \ , 
\label{cubic} \\
H^{(4)}&=&\frac{1}{8N}\text{Tr} (\Psi_0^{-1} \eta \Psi_0^{-1} \eta \Psi_0^{-1} \eta \Psi_0^{-1} \eta \Psi_0^{-1}) \ .
\label{quartic}
\end{eqnarray}
We note that the form of these vertices is the same for both the free (UV) and the interacting (IR) conformal theories (the difference is induced by the different background shifts in the two cases).

Similarly there is a null-plane version of this construction which would correspond to light-cone gauge higher spin theory. This gauge was employed  in \cite{Koch:2010cy} to exhibit the one-to-one map between the two descriptions: the null-plane bi-locals $\Psi(x^+;x_1^-,x_2^-;x_1,x_2)$ and the higher spin fields $\mathcal{H}(x^+;x^-,x,z;\theta)$ in AdS$_4$, where $\theta$ is higher spin coordinate generating the helicities for the  sequence of higher spin fields. On both sides we have the same number of dimensions $1+2+2=1+3+1$, the same representation of the conformal group, and the same number of degrees of freedom. Consequently the mapping that we establish between bi-local and higher spin fields in AdS$_4$ is one-to-one and it involves the extra AdS$_4$ coordinate $z$ in a nontrivial way. As such the collective field construction of the AdS dual avoids holography, all the degrees of freedom are accounted for in the CFT.

The explicit (canonical) transformation given in \cite{Koch:2010cy} reads
\begin{eqnarray}
x^-&=&\frac{x_1^-p_1^++x_2^-p_2^+}{p_1^++p_2^+}  \\
x&=&\frac{x_1 p_1^++x_2 p_2^+}{p_1^++p_2^+}  \\
z&=&\frac{\sqrt{p_1^+ p_2^+}}{p_1^++p_2^+}(x_1-x_2)  \label{formulaz} \\
\theta&=&2\arctan \sqrt{p_2^+ / p_1^+} \ ,
\end{eqnarray}
where $p_i^+$ are the conjugate momenta of $x_i^-$. The map going from the bi-local field to the higher spin field is given by an integral transformation
\begin{eqnarray}
&& \mathcal{H}(x^-,x,z,\theta)=\int dp^+dp^xdp^z \, e^{i(x^-p^++xp^x+zp^z)} \int dp_1^+dp_1dp_2^+dp_2 \cr
&& \delta(p_1^++p_2^+-p^+)\delta(p_1+p_2-p^x)\delta(p_1\sqrt{p_2^+/p_1^+}-p_2\sqrt{p_1^+/p_2^+}-p^z)\cr
&& \delta(2\arctan \sqrt{p_2^+/p_1^+}-\theta)\tilde{\Psi}(p_1^+,p_2^+,p_1,p_2) \ , \label{agreecurrent}
\end{eqnarray}
where $\tilde{\Psi}(p_1^+,p_2^+,p_1,p_2)$ is the Fourier transform of the bi-local field $\Psi(x_1^-,x_2^-,x_1,x_2)$. It was demonstrated in \cite{Koch:2010cy} that under this transformation all the generators of collective field theory map into the generators of light-cone gauge Higher Spin Gravity in the form given by Metsaev \cite{Metsaev:1999ui}.  

We see here an explicit expression for the `extra' AdS coordinate $z$ \eqref{formulaz}, in terms of the relative distance between the bi-local coordinates. As such this formula does have some analogy with the short distance (renormalization group) notion, but is much more specific. An important check regarding the identification of the `extra' AdS coordinate $z$ can be seen by taking the $z \rightarrow 0$ limit. Evaluating the bi-local field at $z=0$ gives after a Fourier expansion  the following ``boundary'' form
\begin{eqnarray}
\mathcal{O}^s = \sum_{k=0}^s\frac{(-1)^k \; \Gamma(s+\frac{1}{2})\Gamma(s+\frac{1}{2})}{k!(s-k)! \; \Gamma(k+\frac{1}{2})\Gamma(s-k+\frac{1}{2})}
(\partial_+)^k \phi \, (\partial_+)^{s-k} \phi \ ,
\end{eqnarray}
which are precisely the conformal composite operators \eqref{agreecurrent} (up to a normalization constant).

As a result, in the bi-local picture one has a clear definition of the AdS boundary at $z=0$ and the notion of boundary amplitudes (boundary $S$-matrix) known from AdS. Due to the construction through collective field theory, one is guaranteed to reproduce the boundary correlators in full agreement with the $O(N)$ model. The bulk/bi-local theory is nonlinear with nonlinearities governed by $1/N=G$ and the correlators are now reproduced in terms of Witten diagrams through higher $n$-point vertices as always in AdS duals. All this provides a nontrivial check of the collective picture and the proposal that bi-local fields provide a bulk representation of AdS$_4$ higher spin fields.



\section{Perturbation Theory: 1/N} \label{sec:CM}

The simplest case of the correspondence involves the UV fixed point CFT of noninteracting $N$-component bosonic or fermionic fields. These theories are characterized by the existence of an infinite sequence of higher spin currents that are conserved. Consequently one has a higher symmetry and an infinite sequence of generators
\begin{eqnarray}
Q^s=\int d\vec{x} J_{0\mu_1\mu_2\cdots \mu_s} \ .
\end{eqnarray}
In such a theory, the Coleman-Mandula theorem would imply that the $S$-matrix should be 1. The relevance and implications of the Coleman-Mandula theorem in AdS$_4$/CFT$_3$ was recently addressed by Maldacena and Zhiboedov \cite{Maldacena:2011jn, Maldacena:2012sf}. They work in the light cone and deduce the consequences of  the infinite sequence of associated charges
\begin{eqnarray}
Q^s=\int dx^-dx J_{s,--\cdots -}
\end{eqnarray}
demonstrating that the (boundary) correlators are necessarily given by free fields, and it is in this sense that the theory can be categorized as simple. 

The recovered correlators $C_n \equiv \langle \mathcal{O}_1 \mathcal{O}_2 \cdots \mathcal{O}_n \rangle$ even though expressible in terms of free $N$-component fields, are nonzero for all $n$. They are, as we have described above, associated with a nonlinear bulk theory, with nonlinearities governed by $1/N=G$. The question then concerns the fate of these nonlinearities characterizing the AdS$_4$ HS theory. In general there is a question regarding the existence of an $S$-matrix in CFT (and also AdS spacetime gravity).

Boundary correlators are sometimes described in the literature as a ``boundary $S$-matrix'' of the AdS theory. Mack \cite{Mack:2009mi, Mack:2009gy} has shown that when written in an integral (Mellin) representation the correlators possess structure equivalent to an $S$-matrix; crossing, duality, etc. Nevertheless, this ``boundary $S$-matrix'' lacks some of the key features of a genuine scattering matrix in particular with regard to equivalence under field redefinitions (central to the discussion of the Coleman-Mandula theorem that we concentrate on next).

Based on the collective construction we are led to consider (and evaluate) another more direct $S$-matrix given by  the physical picture of (collective) dipoles that underlie the CFT$_3$/Higher Spin Holography. In this picture we have an appropriate on-shell relation and a definition of the $S$-matrix through a standard LSZ reduction formula  as a limit of general bi-local correlation functions. In a time-like gauge (single-time), one has the on-shell relation: $E^2-(\vert \vec{k}_1\vert+\vert \vec{k}_2\vert)^2=0$, and the $S$-matrix can be defined by the LSZ-type reduction formula
\begin{eqnarray}
S = \lim \prod_{i} (E_i^2-(\vert \vec{k}_i\vert+\vert \vec{k}_{i'}\vert )^2) \langle \tilde{\Psi}(E_1,\vec{k}_1,\vec{k}_{1'}) \tilde{\Psi}(E_2,\vec{k}_2,\vec{k}_{2'})\cdots \rangle
\label{Sdefintion}
\end{eqnarray}
where the $\tilde{\Psi}$ operators denote energy-momentum transform of the bi-local fields \eqref{bilocal}. The limit implies the on-shell specification for the energies of the dipoles. In the light-cone gauge, (\ref{Sdefintion}) would correspond to
\begin{equation}
\lim \prod_{i} (P_i^--\frac{p_i^2}{2p_i^+}-\frac{p^2_{i'}}{2p^+_{i'}}) \langle \tilde{\Psi}(P_1^-;p_1^+,p_1,p^+_{1'},p_{1'}) \tilde{\Psi} (P_2^-;p^+_2,p_2,p^+_{2'},p_{2'})\cdots \rangle \ .
\end{equation}
We note that the correlation functions appearing in this construction are not the correlation functions of conformal current operators $J_{s,-\cdots-}$. As Maldacena and Zhiboedov have discussed, the Ward identities based on currents provide a reconstruction of correlation functions for bi-local operators of the form $\mathcal{B}(x^+; (x_1^-,x_2^-); x_1=x_2)$. Since these are bi-local in $x$ but local in the other coordinates they would not contain the information for reconstruction of the $S$-matrix.

The details of the evaluation the $S$-matrix are as follows. We have explicitly performed the 3 and 4-point scattering amplitude calculations through the associated Witten diagrams. In momentum space, the relevant cubic (\ref{cubic}) and quartic (\ref{quartic}) interaction vertices take the form
\begin{eqnarray}
H^{(3)} &=& \frac{\sqrt{2}}{\sqrt{N}}\int \prod_{i=1}^3 d\vec{k}_i \bigl[-\frac{\omega_{k_1 k_2 k_3}}{3}\alpha_{\vec{k}_1 \vec{k}_2}\alpha_{-\vec{k}_2 \vec{k}_3}\alpha_{-\vec{k}_3-\vec{k}_1} \cr
&& +\omega_{k_2} \alpha_{\vec{k}_1 \vec{k}_2}\alpha_{-\vec{k}_2 \vec{k}_3}\alpha^\dagger_{\vec{k}_3 \vec{k}_1}+h.c. \bigr] \label{cubicosc} \\
H^{(4)} &=& \frac{1}{N} \int \prod_{i=1}^4 d\vec{k}_i \; \frac{\omega_{k_1 k_2 k_3 k_4}}{4}
\bigl[\alpha_{\vec{k}_1 \vec{k}_2}\alpha_{-\vec{k}_2 \vec{k}_3}\alpha_{-\vec{k}_3 \vec{k}_4}\alpha_{-\vec{k}_4-\vec{k}_1} \cr
&& +4\alpha_{\vec{k}_1 \vec{k}_2}\alpha_{-\vec{k}_2 \vec{k}_3}\alpha_{-\vec{k}_3 \vec{k}_4}\alpha^\dagger_{\vec{k}_4 \vec{k}_1}+h.c. \cr
&& +4\alpha_{\vec{k}_1 \vec{k}_2}\alpha_{-\vec{k}_2 \vec{k}_3}\alpha^\dagger_{\vec{k}_3 \vec{k}_4}\alpha^\dagger_{-\vec{k}_4 \vec{k}_1}
+2\alpha_{\vec{k}_1 \vec{k}_2}\alpha^\dagger_{\vec{k}_2 \vec{k}_3}\alpha_{\vec{k}_3 \vec{k}_4}\alpha^\dagger_{\vec{k}_4 \vec{k}_1} \bigr]
\label{quarticosc}
\end{eqnarray}
where we used the notation $\omega_{k_1k_2\cdots k_i} \equiv \omega_{k_1}+\omega_{k_2}+\cdots+\omega_{k_i}$ and $h.c.$ means taking the hermitian conjugate of {\it only} the terms ahead of it.

\begin{figure}
\begin{center}
\includegraphics[width=0.3\textwidth]{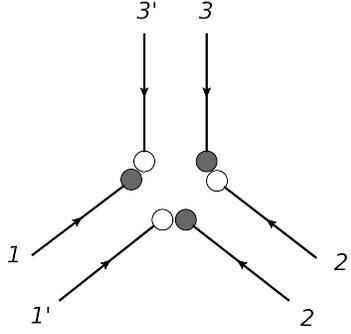}  
\caption{Scattering of three dipoles.}
\label{threescattering}
\end{center}
\end{figure}

For the three-dipole scattering ($1+2 \to 3$), the amplitude is given by
\begin{eqnarray}
\langle 0 \vert \alpha_{\vec{p}_{3} \vec{p}_{3'}} \,
T \exp \left[ -i \int_{-\infty}^\infty dt \, H^{(3)} (t) \right]
\alpha^\dagger_{\vec{p}_{2} \vec{p}_{2'}} \alpha^\dagger_{\vec{p}_{1} \vec{p}_{1'}} \vert 0 \rangle \ ,
\end{eqnarray}
where $T$ means time-ordered. Graphically, this corresponds to the diagram shown in Figure \ref{threescattering}. The evaluation can be easily performed in the {\it interaction} picture giving the result
\begin{eqnarray}
&& \frac{i^3}{8} \delta(E_1+E_2-E_3) [(\omega_{p_{1'}}+\omega_{p_2}) \delta(\vec{p}_1-\vec{p}_3) \cr
&& \qquad \delta( \vec{p}_{2'}-\vec{p}_{3'}) \delta (\vec{p}_{1'}+\vec{p}_2) + \text{7 more terms} ] \cr
&=& \frac{i^3}{8} (E_1+E_2-E_3) \delta(E_1+E_2-E_3) [\delta(\vec{p}_1-\vec{p}_3) \cr
&& \qquad \delta( \vec{p}_{2'}-\vec{p}_{3'}) \delta (\vec{p}_{1'}+\vec{p}_2) + \text{7 more terms} ] 
\end{eqnarray}
where in the second step we have used the energy conservation and the delta functions. The 7 more terms are due to the symmetrization of the propagators over $(1 \leftrightarrow 1', 2 \leftrightarrow 2', 3 \leftrightarrow 3')$. Collecting all the numerical factors, we get the final result 
\begin{align}
S(1+2 & \rightarrow 3)=-\frac{\sqrt{2}}{8\sqrt{N}} (E_1+E_2-E_3) \delta(E_1+E_2-E_3) \cr
& [\delta(\vec{p}_1-\vec{p}_3)\delta(\vec{p}_{2'}-\vec{p}_{3'})\delta(\vec{p}_{1'}+\vec{p}_2)+\text{7 more terms} ] \ .
\end{align}
Clearly the energy pre-factor takes the same form as the conservation term and the result $S_3=0$ follows.

\begin{figure}
\begin{center}
\includegraphics[width=0.8\textwidth]{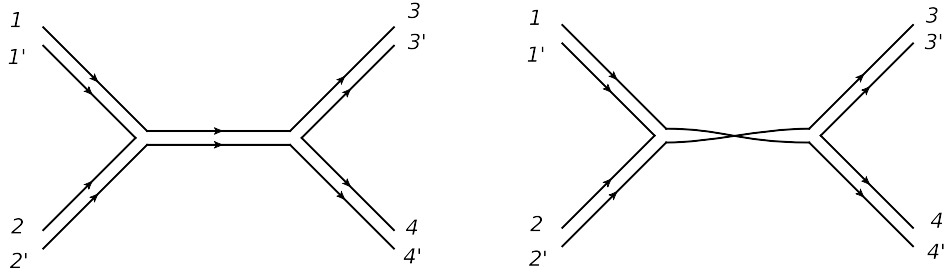}  
\caption{The $s$-channel (and twisted) diagrams of four-dipole scattering.}
\label{fourchanneelsca}
\end{center}
\end{figure}

Next for the four-dipole scattering ($1+2 \to 3+4$), the calculation is similar. The scattering amplitude is given by
\begin{eqnarray}
\langle 0\vert \alpha_{\vec{p}_{3}\vec{p}_{3'}}\alpha_{\vec{p}_{4}\vec{p}_{4'}} \, T \exp \Bigl[
-i\int^\infty_{-\infty} dt \, \bigl( H^{(3)} (t) + H^{(4)} (t) \bigr) \Bigr]
\alpha^\dagger_{\vec{p}_{1} \vec{p}_{1'}}\alpha^\dagger_{\vec{p}_{2} \vec{p}_{2'}}\vert 0 \rangle
\end{eqnarray}
where $H^{(4)}$ is explicitly given in (\ref{quarticosc}). The $1/N$ contributions to the $S_4$ scattering amplitude are
\begin{eqnarray}
&&-\frac{2}{9N}\int  d\vec{k}_i d\vec{l}_j \, \omega_{k_1k_2k_3}\omega_{l_1l_2l_3}\langle 0 \vert \alpha_{\vec{p}_{3} \vec{p}_{3'}}\alpha_{\vec{p}_{4} \vec{p}_{4'}}\alpha_{\vec{k}_1 \vec{k}_2}\alpha_{-\vec{k}_2 \vec{k}_3}\alpha_{-\vec{k}_3-\vec{k}_1} \cr
&& \qquad \qquad \alpha^\dagger_{\vec{l}_1 \vec{l}_2}\alpha^\dagger_{-\vec{l}_2 \vec{l}_3}\alpha^\dagger_{-\vec{l}_3-\vec{l}_1}\alpha^\dagger_{\vec{p}_{1} \vec{p}_{1'}}\alpha^\dagger_{\vec{p}_{2} \vec{p}_{2'}}\vert 0 \rangle \cr
&&-\frac{2}{N}\int  d\vec{k}_i d\vec{l}_j \, \omega_{k_2}\omega_{l_2}\langle 0 \vert \alpha_{\vec{p}_{3} \vec{p}_{3'}}\alpha_{\vec{p}_{4} \vec{p}_{4'}}\alpha_{\vec{k}_1 \vec{k}_2}\alpha_{-\vec{k}_2 \vec{k}_3}\alpha^\dagger_{\vec{k}_3 \vec{k}_1} \cr
&& \qquad \qquad \alpha^\dagger_{\vec{l}_1 \vec{l}_2}\alpha^\dagger_{-\vec{l}_2 \vec{l}_3}\alpha_{\vec{l}_3 \vec{l}_1}\alpha^\dagger_{\vec{p}_{1} \vec{p}_{1'}}\alpha^\dagger_{\vec{p}_{2} \vec{p}_{2'}}\vert 0 \rangle \cr
&&-\frac{i}{N}\int  d\vec{k}_i \, \omega_{k_1k_2k_3k_4}\langle 0\vert \alpha_{\vec{p}_{3} \vec{p}_{3'}}\alpha_{\vec{p}_{4} \vec{p}_{4'}}\alpha_{\vec{k}_1 \vec{k}_2}\alpha_{-\vec{k}_2 \vec{k}_3} \cr
&& \qquad \qquad \alpha^\dagger_{\vec{k}_3 \vec{k}_4}\alpha^\dagger_{-\vec{k}_4 \vec{k}_1}\alpha^\dagger_{\vec{p}_{1} \vec{p}_{1'}}\alpha^\dagger_{\vec{p}_{2} \vec{p}_{2'}}\vert 0 \rangle \cr
&&-\frac{i}{2N}\int  d\vec{k}_i \, \omega_{k_1k_2k_3k_4}\langle 0 \vert \alpha_{\vec{p}_{3} \vec{p}_{3'}}\alpha_{\vec{p}_{4} \vec{p}_{4'}}\alpha_{\vec{k}_1 \vec{k}_2}\alpha^\dagger_{\vec{k}_2 \vec{k}_3} \cr
&& \qquad \qquad \alpha_{\vec{k}_3 \vec{k}_4}\alpha^\dagger_{\vec{k}_4 \vec{k}_1}\alpha^\dagger_{\vec{p}_{1} \vec{p}_{1'}}\alpha^\dagger_{\vec{p}_{2} \vec{p}_{2'}}\vert 0 \rangle \ .
\label{fourlines}
\end{eqnarray}
The first term in (\ref{fourlines}) has only $s$-channel contributions shown in Figure \ref{fourchanneelsca}, while the second term in (\ref{fourlines}) has all $s,t,u$-channel contributions. The $s$-channel diagrams and their twisted ones (due to the symmetrization of propagators) are summed to be
\begin{eqnarray}
&& \frac{i}{8 N} \delta(E_1+E_2-E_3-E_4) \bigl[ (\omega_{p_{2'}}+\omega_{p_3}) \delta(\vec{p}_1-\vec{p}_3) \delta(\vec{p}_{1'}+\vec{p}_2) \cr
&& \qquad \qquad \delta( \vec{p}_{2'}-\vec{p}_{4'}) \delta (\vec{p}_{3'}+\vec{p}_4) + \text{15 more terms} \cr
&& \qquad \qquad +(\omega_{p_{1'}}+\omega_{p_{3}}) \delta(\vec{p}_2-\vec{p}_3) \delta(\vec{p}_1+\vec{p}_{2'}) \cr
&& \qquad \qquad \delta( \vec{p}_{1'}-\vec{p}_{4'}) \delta (\vec{p}_{3'}+\vec{p}_4) + \text{15 more terms} \bigr] \ .
\end{eqnarray}

The $t,u$-channel diagrams together, with their twisted diagrams, are calculated in a similar way. In addition one also has the cross-shaped diagram 
coming from the four point vertex. We omit the details of the evaluation, the reader can consult \cite{deMelloKoch:2012vc}.

Summing all the diagrams, the final result for the four-dipole scattering is found to be
\begin{align}
& S(1+2\rightarrow 3+4)= \frac{i}{8 N} \delta(E_1+E_2-E_3-E_4) \cr
& \quad \times \bigl[ (\omega_{p_2}-\omega_{p_4}) \delta(\vec{p}_1-\vec{p}_3) \delta(\vec{p}_{1'}+\vec{p}_2) \delta( \vec{p}_{2'}-\vec{p}_{4'}) \delta (\vec{p}_{3'}+\vec{p}_4) + \text{15 more terms} \cr
& \quad +(\omega_{p_{2'}}-\omega_{p_{3'}}) \delta(\vec{p}_2-\vec{p}_3) \delta(\vec{p}_1+\vec{p}_{2'}) \delta( \vec{p}_{1'}-\vec{p}_{4'}) \delta (\vec{p}_{3'}+\vec{p}_4) + \text{15 more terms} \bigr] \cr
&= \frac{i}{16 N} (E_1+E_2-E_3-E_4) \delta(E_1+E_2-E_3-E_4)  \cr
& \quad \times \bigl[ \delta(\vec{p}_1-\vec{p}_3) \delta(\vec{p}_{1'}+\vec{p}_2) \delta( \vec{p}_{2'}-\vec{p}_{4'}) \delta (\vec{p}_{3'}+\vec{p}_4) + \text{15 more terms} \cr
& \quad +\delta(\vec{p}_2-\vec{p}_3) \delta(\vec{p}_1+\vec{p}_{2'}) \delta( \vec{p}_{1'}-\vec{p}_{4'}) \delta (\vec{p}_{3'}+\vec{p}_4) + \text{15 more terms} \bigr] 
\end{align}
which now implies that $S_4=0$.

Based on this examples  one can  expect that the $S$-matrix equals $1$ for the bi-local theory of the free UV fixed point. We emphasize that this is a statement about the bulk theory where the $1/N$ vertices are nonzero even for the duality based on the free CFT (the nonzero vertices are required to reproduce arbitrary $n$-point correlators through Witten-Feynman diagrams). Since we view the collective representation as a particular gauge fixed description of Vasiliev's HS theory, we suggest that analogous statement holds there in any gauge. Our definition of the $S$-matrix applies to AdS, it is implemented through a simple Fourier transform map to the dipole picture. It is clear that any change of boundary conditions will result in a non-trivial $S$-matrix, since the cancellation that we saw above will not take place. We also mention that in the framework of BCFW recursions for Higher Spin interactions, the relevance of extended observables was noted in \cite{Fotopoulos:2010ay, Dempster:2012vw}. Finally the field theoretic equivalence principle would now imply that there should be a field transformation which removes the $G=1/N$ interactions of the bulk/bi-local theory. We were able to establish the existence of such a field redefinition in exact terms. It will follow as an easy corollary of a more general, geometric formulation of the bi-local theory that we describe next.



\section{Geometric Formulation} \label{sec:Field}

It is useful to describe the bi-local/bulk theory in an equivalent geometric scheme. This is based on the algebra of all bi-local observables of the $N$-component bosonic (or fermionic) CFT which close an algebra at large $N$ constrained by the corresponding Casimir operator. One therefore has an algebraic pseudo-spin system whose nonlinearity is governed by the coupling constant $G=1/N$. As such they have been employed earlier for developing a large $N$ expansion \cite{Jevicki:1980mj} and as a model for quantization \cite{Berezin:1978sn}. The pseudo-spin variables can be seen to obey a closed set of equations equivalent to the collective formulation. This version of the theory is in its perturbative ($1/N$) expansion identical to the bi-local collective representation. But the algebraic pseudo-spin representation has certain advantages, especially for studies at the exact non-perturbative level. In addition this construction is equally simple implemented for the case of $N$-component fermions (or anti-commuting) Grassmann fields as it is for $N$-component bosonic fields. Being algebraic in nature the only difference between turns out to be in the compact versus the non-compact nature of the pseudo-spin algebra. This will be seen to correspond to a change of $N$ to $-N$. Consequently it will follow from the geometric pseudo-spin representation that there is a simple perturbative correspondence between the two system, namely the one of changing the coupling constant $G$ to the opposite $-G$. This feature is the characteristic of continuation from Anti de Sitter to de Sitter spacetime. The second feature of the algebraic formulation is that it will provide a natural scheme for quantization. In particular the Hilbert space at non-perturbative level will be given through geometric (K\"ahler) quantization. It will be seen to faithfully represent the quantum theory at finite $N$.

For the bosonic $O(N)$ system the (pseudo) spin bi-local operators are given by
\begin{eqnarray}
S(\vec{p}_1,\vec{p}_2)&=&\frac{1}{\sqrt{2N}}\sum_i a^i(\vec{p}_1)a^i(\vec{p}_2) \ , \\
S^\dagger(\vec{p}_1,\vec{p}_2)&=&\frac{1}{\sqrt{2N}}\sum_i a^{i\, \dagger}(\vec{p}_1)a^{i\, \dagger}(\vec{p}_2) \ , \\
B(\vec{p}_1,\vec{p}_2)&=&\sum_i a^{i\,\dagger}(\vec{p}_1)a^i(\vec{p}_2) \ . 
\end{eqnarray}
where we have used creation-annihilation operators of the component field. The Hamiltonian is essentially equal to $B$ (for the free case for example)
\begin{eqnarray}
H=\int d \vec{p} \; {\cal H}(\vec{p},\vec{p}) \ , \qquad
{\cal H}(\vec{p},\vec{p}) = \omega_{\vec{p}} \, B(\vec{p},\vec{p})
+ \frac{N}{2}\omega_{\vec{p}} \, \delta(\vec{0}) \ .
\label{Hamiltonian}
\end{eqnarray}
The bi-local pseudo-spin variables close an algebra
\begin{eqnarray}
\big[ S(\vec{p}_1,\vec{p}_2),S^\dagger(\vec{p}_3,\vec{p}_4)\big] &=& 
\frac{1}{2}\left(\delta_{\vec{p}_2,\vec{p}_3}\delta_{\vec{p}_4,\vec{p}_1}+\delta_{\vec{p}_2,\vec{p}_4}\delta_{\vec{p}_3,\vec{p}_1}\right) \cr
&& +\frac{1}{2 N}\bigl[\delta_{\vec{p}_2,\vec{p}_3}B(\vec{p}_4,\vec{p}_1)+\delta_{\vec{p}_2,\vec{p}_4}B(\vec{p}_3,\vec{p}_1) \cr
&& +\delta_{\vec{p}_1,\vec{p}_3}B(\vec{p}_4,\vec{p}_2)+\delta_{\vec{p}_1,\vec{p}_4}B(\vec{p}_3,\vec{p}_2) \bigr] \ , \\
\big[ B(\vec{p}_1,\vec{p}_2),S^\dagger(\vec{p}_3,\vec{p}_4)\big] &=&
\delta_{\vec{p}_2,\vec{p}_3}S^\dagger(\vec{p}_1,\vec{p}_4)
+\delta_{\vec{p}_2,\vec{p}_4}S^\dagger(\vec{p}_1,\vec{p}_3) \ , \\
\big[ B(\vec{p}_1,\vec{p}_2),S(\vec{p}_3,\vec{p}_4)\big] &=&
-\delta_{\vec{p}_1,\vec{p}_3}S(\vec{p}_2,\vec{p}_4)
-\delta_{\vec{p}_1,\vec{p}_4}S(\vec{p}_2,\vec{p}_3) \ .
\end{eqnarray}
In the O(N) singlet sector one finds that  the quadratic (Casimir) operator associated with the algebra is constrained to equal
\begin{eqnarray}
-\frac{8}{N}S^\dagger\star S +\left(1+\frac{2}{N}B\right)\star \left(1+\frac{2}{N}B\right) = \mathbb{I} \ ,
\end{eqnarray}
where we have  the matrix star product notation: $A \star B=\int d\vec{p}_2 A(\vec{p}_1 , \vec{p}_2)B(\vec{p}_2 , \vec{p}_3)$. The importance of the Casimir constraint is that it implies that the above non-commuting set of bi-local operators is not independent. In particular it implies that the bi-local pseudo-spin algebra has representations in terms of a canonical pairs of bi-local variables. The canonical collective theory based provides a specific representation of the above algebra. Explicitly, one has
\begin{eqnarray}
&& S(\vec{x}_1,\vec{x_2})=\int d\vec{p}_1d\vec{p}_2d\vec{y}_1d\vec{y}_2e^{i\vec{p}_1\cdot(\vec{x}_1-\vec{y}_1)}e^{i\vec{p}_2\cdot(\vec{x}_2-\vec{y}_2)} \cr
&& \Bigl[\frac{-2}{\sqrt{\omega_{p_1} \omega_{p_2}}}\Pi(\vec{y}_1,\vec{z}_1)\star\Psi(\vec{z}_1,\vec{z}_2)\star\Pi(\vec{z}_2,\vec{y}_2) \cr
&& -i\sqrt{N}\sqrt{\frac{\omega_{p_2}}{\omega_{p_1}}}\Psi(\vec{y}_2,\vec{z}_1)\star\Pi(\vec{y}_1,\vec{z}_1)
-i\sqrt{N}\sqrt{\frac{\omega_{p_1}}{\omega_{p_2}}}\Psi(\vec{y}_1,\vec{z}_1)\star\Pi(\vec{y}_2,\vec{z}_1) \cr
&& -\frac{N}{8}\frac{1}{\sqrt{\omega_{p_1} \omega_{p_2}}} \Psi^{-1}(\vec{y}_1,\vec{y}_2)
+\frac{N\sqrt{\omega_{p_1} \omega_{p_2}}}{2}\Psi(\vec{y}_1,\vec{y}_2) \Bigr]
\end{eqnarray}
and a similar representation for $B$. Transforming them into momentum space and expanding we generate the $1/N$ series
\begin{eqnarray}
&& S(\vec{k}_1,\vec{k}_2)=\alpha_{\vec{k}_1\vec{k}_2}-\frac{1}{\sqrt{2N}}\Bigl[
\alpha_{\vec{k}_1\vec{k}_3}\star\alpha_{-\vec{k}_3\vec{k}_2}
-\alpha_{\vec{k}_1\vec{k}_3}^\dagger\star\alpha_{-\vec{k}_3\vec{k}_2}^\dagger \cr
&& \qquad\qquad~ -\alpha_{\vec{k}_1\vec{k}_3}\star\alpha_{\vec{k}_3\vec{k}_2}^\dagger
-\alpha_{\vec{k}_1\vec{k}_3}\star\alpha_{\vec{k}_3\vec{k}_2}^\dagger\Bigr]+O(\alpha^3) \ , \\
&& B(\vec{k}_1,\vec{k}_2)=\alpha_{\vec{k}_1\vec{k}_3}\star\alpha_{\vec{k}_3\vec{k}_2}^\dagger
+\alpha_{\vec{k}_1\vec{k}_3}^\dagger\star\alpha_{\vec{k}_3\vec{k}_2} \cr
&& \qquad\qquad~ +2\sqrt{\frac{2}{N}}\Bigl[\alpha_{\vec{k}_1\vec{k}_3}\star\alpha_{-\vec{k}_3\vec{k}_4}
\star\alpha_{-\vec{k}_4\vec{k}_2} \cr
&& \qquad\qquad~ +\alpha_{\vec{k}_1\vec{k}_3} \star\alpha_{\vec{k}_3\vec{k}_4}^\dagger\star\alpha_{\vec{k}_4\vec{k}_2}
-\alpha_{\vec{k}_1\vec{k}_3}^\dagger\star\alpha_{\vec{k}_3\vec{k}_4}\star\alpha_{\vec{k}_4\vec{k}_2}^\dagger \cr
&& \qquad\qquad~ -\alpha_{\vec{k}_1\vec{k}_3}^\dagger\star\alpha_{-\vec{k}_3\vec{k}_4}^\dagger
\star\alpha_{-\vec{k}_4\vec{k}_2}^\dagger \Bigr]+O(\alpha^4) \ . 
\end{eqnarray}

For the collective field theory this gives the $1/N$ vertices of the nonlinear Hamiltonian and also the perturbation expansion for its exact eigenstates. We can now explain how the algebraic formulation is used to establish the linearizing field transformation that we have alluded to earlier.   

The key to the argument is the fact that one can write another realization of the pseudo-spin algebra obeying the Casimir constraint, a realization in which the Hamiltonian is quadratic. It is given by a new canonical set of oscillators $\beta (\vec{p}_1,\vec{p}_2)$ defined by
\begin{eqnarray}
\beta(\vec{p}_1,\vec{p}_2) &=& \Bigl( 1+\frac{1}{N} B \Bigr)^{-\frac{1}{2}}(\vec{p}_1,\vec{p})
\star S(\vec{p},\vec{p}_2) \label{invbetabilocalrep} \\
\beta^\dagger(\vec{p}_1,\vec{p}_2) &=& S^\dagger(\vec{p}_1,\vec{p})
\star \Bigl(1+\frac{1}{N}B \Bigr)^{-\frac{1}{2}}(\vec{p},\vec{p}_2)
\end{eqnarray}
such that
\begin{eqnarray}
B(\vec{p}_1,\vec{p}_2) = \beta^\dagger(\vec{p}_1,\vec{p})\star \beta(\vec{p},\vec{p}_2) \\
\left[\beta (\vec{p}_1,\vec{p}_2),\beta^\dagger (\vec{p}_3,\vec{p}_4)\right] = \delta_{\vec{p}_1,\vec{p}_4}\delta_{\vec{p}_2,\vec{p}_3} \ .
\end{eqnarray}
We see that in this realization the Hamiltonian is quadratic due to \eqref{Hamiltonian}. This therefore specifies the field transformation needed in exact terms.

One can work out the transformation between the fields explicitly, using \eqref{invbetabilocalrep} and expanding in $1/N$  we obtain
\begin{eqnarray}
&& \beta(\vec{k}_1,\vec{k}_2)=\alpha_{\vec{k}_1\vec{k}_2}-\frac{1}{\sqrt{2N}}\Bigl[
\alpha_{\vec{k}_1\vec{k}_3}\star\alpha_{-\vec{k}_3\vec{k}_2}
-\alpha_{\vec{k}_1\vec{k}_3}^\dagger\star\alpha_{-\vec{k}_3\vec{k}_2}^\dagger \cr
&& \qquad\qquad~ -\alpha_{\vec{k}_1\vec{k}_3}\star\alpha_{\vec{k}_3\vec{k}_2}^\dagger -\alpha_{\vec{k}_1\vec{k}_3}\star\alpha_{\vec{k}_3\vec{k}_2}^\dagger \Bigr]+O(\alpha^3) \ ,
\end{eqnarray}
which can be checked to represent a canonical transformation (in the sense of Poisson brackets). We have therefore a construction of the field transformation (in bi-local space) that linearizes the nonlinear $1/N$ Hamiltonian. This was specified at the planar level, namely the tree approximation but can be easily extended (along the lines given above) to all loops. Under this transformation the correlation functions change but the $S$-matrix does not. This is the implementation of the Coleman-Mandula theorem in the large $N$ dual associated with the free field CFT.

\subsection{$Sp(2N)$ Fermions / de Sitter Duality}

We have already emphasized that the pseudo-spin description applies equally to fermionic fields. Among others (corresponding to critical Gross-Neveu type theories) one also has the $Sp(2N)$ CFT that potentially corresponds to de Sitter space duality \cite{Anninos:2011ui}. The key to the de Sitter correspondence is the property that correlation functions between AdS and dS Higher Spin theories correspond simply to a change of $G$ to $-G$. This property is seen to arise naturally in the pseudo-spin formulation. As a consequence the map established for AdS will apply (through analytic continuation) to dS$_4$ Higher Spin at the perturbative level. We now describe the bi-local construction for the $Sp(2N)/dS$ system following \cite{Das:2012dt}.

The theory in question is based on $Sp(2N)$ fermions \cite{LeClair:2007iy} with the action
\begin{equation}
S=\int d\vec{x} \, dt \, (\partial^\mu \eta^i_1 \, \partial_\mu \eta^i_2) \ , \qquad i=1,\ldots,N
\end{equation}
producing the canonical anti-commutation relations
\begin{equation}
\{\eta^i_1(\vec{x},t) , \partial_t \eta^j_2(\vec{x}',t)\}=-\{\eta^i_2(\vec{x},t) , \partial_t \eta^j_1(\vec{x}',t)\} = i\delta(\vec{x}-\vec{x}')\delta^{ij} \ .
\end{equation}
The quantization following \cite{LeClair:2007iy} is based on the mode expansion
\begin{eqnarray}
\eta^i_1(\vec{x} , t)&=&\int \frac{d\vec{k}}{\sqrt{2\omega_k}}(a_{\vec{k}+}^{i\dagger} e^{-i k \cdot x}+a^i_{\vec{k}-}e^{i k \cdot x}) \\
\eta^i_2(\vec{x} , t)&=&\int \frac{d\vec{k}}{\sqrt{2\omega_k}}(-a_{\vec{k}-}^{i\dagger} e^{-i k \cdot x}+a^i_{\vec{k}+}e^{i k \cdot x})
\end{eqnarray}
where $k \cdot x = \omega_k t - \vec{k} \cdot \vec{x}$ and the anti-commutators
\begin{eqnarray}
\{a^i_{\vec{k}-},a_{\vec{k}'-}^{j\dagger}\}=\{a^i_{\vec{k}+},a_{\vec{k}'+}^{j\dagger}\}=\delta(\vec{k}-\vec{k}')\delta^{ij} \ .
\end{eqnarray}
Here the operators $\eta^i_a$ are not hermitian, but pseudo-hermitian, this is a theory with modified hermiticity rules.

Pseudo-spin bi-local variables are introduced based on $Sp(2N)$ invariance, we have a complete set of $Sp(2N)$ invariant operators as follows
\begin{eqnarray}
S(\vec{p}_1,\vec{p}_2)&=&\frac{i}{2\sqrt{N}}\sum_{i=1}^N \Bigl(a_{\vec{p}_1 +}^i a_{\vec{p}_2 -}^i + a_{\vec{p}_2 +}^i a_{\vec{p}_1 -}^i \Bigr) \\
S^\dagger(\vec{p}_1,\vec{p}_2)&=&\frac{i}{2\sqrt{N}}\sum_{i=1}^N \Bigl(a_{\vec{p}_1 +}^{i\dagger} a_{\vec{p}_2 -}^{i\dagger}+a_{\vec{p}_2 +}^{i\dagger} a_{\vec{p}_1 -}^{i\dagger} \Bigr) \\
B(\vec{p}_1,\vec{p}_2)&=&\sum_{i=1}^N a_{\vec{p}_1 +}^{i\dagger} a_{\vec{p}_2 +}^{i} + a_{\vec{p}_1 -}^{i\dagger} a_{\vec{p}_2 -}^{i} \ .
\end{eqnarray}
These invariant operators close an invariant algebra, with the commutation relations given by
\begin{eqnarray}
\big[ S(\vec{p}_1,\vec{p}_2), S^\dagger(\vec{p}_3,\vec{p}_4)\big] &=& \frac{1}{2} \left(
\delta_{\vec{p}_2,\vec{p}_3}\delta_{\vec{p}_4,\vec{p}_1} + \delta_{\vec{p}_2,\vec{p}_4}\delta_{\vec{p}_3,\vec{p}_1}\right) \cr
&&-\frac{1}{4N}\left[\delta_{\vec{p}_2,\vec{p}_3}B(\vec{p}_4,\vec{p}_1)+\delta_{\vec{p}_2,\vec{p}_4}B(\vec{p}_3,\vec{p}_1)\right. \cr
&&+\left. \delta_{\vec{p}_1,\vec{p}_3}B(\vec{p}_4,\vec{p}_2)+
\delta_{\vec{p}_1,\vec{p}_4}B(\vec{p}_3,\vec{p}_2)\right] \\
\big[ B(\vec{p}_1,\vec{p}_2),S^\dagger(\vec{p}_3,\vec{p}_4)\big]&=&
\delta_{\vec{p}_2,\vec{p}_3}S^\dagger(\vec{p}_1,\vec{p}_4)+
\delta_{\vec{p}_2,\vec{p}_4}S^\dagger(\vec{p}_1,\vec{p}_3) \\
\big[ B(\vec{p}_1,\vec{p}_2),S(\vec{p}_3,\vec{p}_4)\big]&=&-
\delta_{\vec{p}_1,\vec{p}_3}S(\vec{p}_2,\vec{p}_4)-
\delta_{\vec{p}_1,\vec{p}_4}S(\vec{p}_2,\vec{p}_3) \ .
\end{eqnarray}
The constraint involving the Casimir operator of the algebra can be shown to take the form
\begin{eqnarray}
\frac{4}{N}S^\dagger\star S+(1-\frac{1}{N}B)\star (1-\frac{1}{N}B)=\mathbb{I} \ .
\label{Casimir}
\end{eqnarray}
The form of the Casimir, which commutes with the above pseudo-spin fields points to the compact nature of the bi-local pseudo-spin algebra associated with the $Sp(2N)$ theory. Indeed comparing with the algebra and Casimir found in the bosonic case, one observes a change of sign.

We can therefore see that the singlet sectors of the fermionic $Sp(2N)$ theory and the bosonic $O(2N)$ theory can be described in a unified algebraic formulations with a quadratic Casimir constraint of the form
\begin{eqnarray}
4\gamma S^\dagger\star S+(1-\gamma B)\star (1-\gamma B) =\text{$\mathbb{I}$}
\end{eqnarray}
where the coupling constant $\gamma=\frac{1}{N}(-\frac{1}{N})$ for the fermionic (bosonic) case respectively. This signifies the compact versus the non-compact nature of the algebra, but also exhibits the relationship obtained through the $N \leftrightarrow -N$ switch that was central in the argument for de Sitter correspondence in \cite{Anninos:2011ui}.

From this algebraic bi-local formulation one can easily see the the collective field representation in both cases. Very simply, the Casimir constraints can be solved, and the algebra implemented in terms of a canonical pair of bi-local fields
\begin{eqnarray}
&& S(\vec{p}_1,\vec{p}_2) = \frac{\sqrt{-\gamma}}{2}\int d\vec{y}_1 d\vec{y}_2 e^{-i(\vec{p}_1 \cdot \vec{y}_1+\vec{p}_2 \cdot \vec{y}_2)} \cr
&& \qquad \times \Bigl\{-\frac{2}{\kappa_{p_1}\kappa_{p_2}}\Pi\star\Psi\star\Pi(\vec{y}_1,\vec{y}_2)
-\frac{1}{2\gamma^2\kappa_{p_1}\kappa_{p_2}} \Psi^{-1}(\vec{y}_1,\vec{y}_2) \cr
&& \qquad +\frac{\kappa_{p_1}\kappa_{p_2}}{2}\Psi(\vec{y}_1,\vec{y}_2)
-i\frac{\kappa_{p_1}}{\kappa_{p_2}}\Psi\star\Pi(\vec{y}_1,\vec{y}_2)
-i\frac{\kappa_{p_2}}{\kappa_{p_1}}\Pi\star\Psi(\vec{y}_1,\vec{y}_2) \Bigr\} \\
&& B(\vec{p}_1,\vec{p}_2) = \frac{1}{\gamma}+\int d\vec{y}_1 d\vec{y}_2 e^{-i(\vec{p}_1 \cdot \vec{y}_1+\vec{p}_2 \cdot \vec{y}_2)} \cr
&& \qquad \times \Bigl\{\frac{2}{\kappa_{p_1}\kappa_{p_2}}\Pi\star\Psi\star\Pi(\vec{y}_1,\vec{y}_2)
+\frac{1}{2\gamma^2\kappa_{p_1}\kappa_{p_2}} \Psi^{-1}(\vec{y}_1,\vec{y}_2) \cr
&& \qquad +\frac{\kappa_{p_1}\kappa_{p_2}}{2}\Psi(\vec{y}_1,\vec{y}_2)
-i\frac{\kappa_{p_1}}{\kappa_{p_2}}\Psi\star\Pi(\vec{y}_1,\vec{y}_2)
+i\frac{\kappa_{p_2}}{\kappa_{p_1}}\Pi\star\Psi(\vec{y}_1,\vec{y}_2) \Bigr\}
\end{eqnarray}
where $\kappa_{p} \equiv \sqrt{\omega_{p}}$. 

Recalling that the Hamiltonian is given in terms of $B$ we now see that its bi-local form is the same in the fermionic and the bosonic case. While the bi-local field representation of $B$ is the same in the fermionic and bosonic cases, the difference is seen in the representations of operators $S$ and $S^\dagger$. These operators create singlet states in the Hilbert space and the difference contained in the sign of gamma implies the opposite shifts for the background fields. The algebraic pseudo spin reformulation is therefore seen to account for all the perturbative ($1/N$) features of the the bi-local theory. However, in addition, the algebraic formulation will provide a proper framework for quantization and for defining the bi-local Hilbert space.

\subsection{Geometric Quantization and the Hilbert Space}

The bi-local pseudo-spin algebra has several equivalent representations that turn out to be useful. Beside that collective representation that we have explained above, one has an equivalent geometric (K\"ahler space) description. It will be seen capable of incorporating non-perturbative features related to the Grassmanian origin of bi-local fields and its Hilbert space. One issue which is usually raised in the use of collective bi-local variables is their over completeness. Namely, at any finite $N$ these set of variables over counts the original set given by the $N$-component local field. As such quantization in the bi-local space is suspect of grossly over counting the total number of states in the Hilbert space. This would prevent the use of this scheme in any non-perturbative (and finite $N$) context. One way the over completeness issue can be resolved is by introducing constraints which correctly reduce the number of variables. This procedure even though correct in principle is difficult to implement. Another, much more elegant scheme, which will be featured below is geometric quantization. This as we explain will lead to exact finite $N$ Hilbert space.

The K\"ahler parameterization of the pseudo-spins is given by
\begin{eqnarray}
S&=&Z\star(1+\frac{1}{N}\bar{Z}\star Z)^{-1} \\
S^\dagger &=&(1+\frac{1}{N}\bar{Z}\star Z)^{-1}\star \bar{Z} \\
B &=&2\;Z\star (1+\frac{1}{N}\bar{Z} \star Z)^{-1}\star \bar{Z} \ .
\end{eqnarray}
 One can in this representation write down a Lagrangian for the bi-local K\"ahler theory
\begin{eqnarray}
\mathcal{L}=i\int dt \, \text{Tr}[Z(1+\frac{1}{N}\bar{Z}Z)^{-1}\dot{\bar{Z}}-\dot{Z}(1+\frac{1}{N}\bar{Z}Z)^{-1}\bar{Z}]-\mathcal{H} \ .
\end{eqnarray}
For regularization purposes, it is useful to consider putting $\vec{x}$ in a box and limiting the momenta by a cutoff $\Lambda$; this makes the bi-local fields into finite dimensional matrices (which we will take to be a size $K$). For $Sp(2N)$ one deals with a $K \times K$ dimensional complex matrix $Z$ and we have obtained in the above a compact symmetric (K\"ahler) space
\begin{eqnarray}
ds^2=\text{tr}[dZ(1-\bar{Z}Z)^{-1}d\bar{Z}(1-Z\bar{Z})^{-1}] \ .
\end{eqnarray}
Quantization on K\"ahler manifolds in general has been formulated in detail by Berezin \cite{Berezin:1975}. According to the classification of \cite{Berezin:1975}, the above corresponds to the manifold labelled by $M_I(K,K)$. We note that the standard fermionic problem corresponds to another manifold $M_{III}(K,K)$ of complex antisymmetric matrices. We will now summarize some of the results of this quantization which are directly relevant to the bi-local collective field theory. Commutation relations of this system follow from the Poisson brackets associated with the Lagrangian $\mathcal{L}(\bar{Z},Z)$. States in the Hilbert space are represented by (holomorphic) functions (functionals) of the bi-locals $Z(k,l)$. A K\"ahler scalar product defining the bi-local Hilbert space reads
\begin{equation}
(F_1,F_2)=C(N,K)\int d\mu(\bar{Z},Z)F_1(Z)F_2(\bar{Z})\det [1+\bar{Z}Z]^{-N}
\end{equation}
with the (Kahler) integration measure
\begin{eqnarray}
d\mu=\det[1+\bar{Z}Z]^{-2K}d\bar{Z}dZ \ .
\end{eqnarray}
The normalization constant is found from requiring $(F_1,F_1)=1$ for $F_1=1$ which leads to
\begin{eqnarray}
a(N,K)=\frac{1}{C(N,K)}=\int d\mu(\bar{Z},Z)\det [1+\bar{Z}Z]^{-N} \ .
\end{eqnarray}
This is matrix (complex Penner Model) integral determining $C(N,K)$.

Based on this representation one can address the question concerning the number of states in the above Hilbert space. As mentioned earlier naively, bi-local theory would seem to over-count the number of degrees of freedom due to the fact that (at finite $N$) bi-locals represent an over-complete set of dynamical variables. Originally one essentially had $2NK$ fermionic degrees of freedom with a finite Hilbert space. The bi-local description is based on (complex) bosonic variables of dimensions $K^2$ and the corresponding Hilbert space would appear to be much larger. 

In geometric (K\"ahler) quantization however due to the compact nature of the phase space, the number of states turns out to be cutoff by a non-perturbative effect proportional to $N$. In \cite{Das:2012dt} an evaluation of the number of states in the geometric Hilbert space (at finite $N$ and $K$) for the $Sp(2N)$ system was given showing that the dimension of the bi-local Hilbert space in geometric (K\"ahler) quantization agrees with the dimension of the singlet Hilbert space of the $Sp(2N)$ fermionic theory. Elements of this estimate go as follows.

The dimension of quantized Hilbert space is found from considering the trace of the unit operator $\hat{O}=I$
\begin{equation}
\text{Tr}(I)=C(N,K) \int \prod_{k,l=1}^K d\bar{Z}(k,l)dZ(k,l)\det [1+\bar{Z}Z]^{-2K} \ .
\end{equation}
Consequently the dimension of the bi-local Hilbert space is given by
\begin{eqnarray}
\dim \mathcal{H}_B=\frac{C(N,K)}{C(0,K)}=\frac{a(0,K)}{a(N,K)} \ .
\end{eqnarray}
This matrix (Penner type) integral can be evaluated in several steps \cite{Das:2012dt, Berezin:1978sn}. First a diagonalization of the (complex) matrix gives the integration measure 
\begin{eqnarray}
[d\bar{Z}dZ]=\vert \Delta(\omega) \vert^2 \prod_{l=1}^K d\omega_l d\Omega
\end{eqnarray}
where $d\Omega$ and  $d\omega$ denote ``angular'' and diagonal components of the matrix and $\Delta(x_1,\cdots,x_K)=\prod_{k<l}(x_k-x_l)$ is a Van der Monde determinant, with $x_i=\omega_i^2$. 
Consequently the matrix integral for $a(N,K)$ (and $C(N,K)$) becomes
\begin{eqnarray}
a(N,K)=\frac{\text{Vol}(\Omega)}{K!}\int \Delta(x_1,\cdots,x_K)^2 \prod_l (1+\omega_l^2)^{-2K-N}\prod_l d\omega_l \ .
\end{eqnarray}
This integral can be evaluated exactly \cite{Das:2012dt}, resulting in the following formula for the number of states in the bi-local $Sp(2N)$ Hilbert space
\begin{eqnarray}
\dim \mathcal{H}_B=\prod_{j=0}^{K-1}\frac{\Gamma(j+1)\Gamma(N+K+j+1)}{\Gamma(K+j+1)\Gamma(N+j+1)} \ .
\label{counting}
\end{eqnarray}
This number can be compared with explicit enumeration of $Sp(2N)$ invariant states in the fermionic Hilbert space (for low values of $N$ and $K$) with complete agreement. This settles however the potential problem of over-completeness of the bi-local representation. Since the $Sp(2N)$ counting uses the fermionic nature of creation operators and features exclusion when occupation numbers grow above certain limit it is seen that bi-local geometric quantization elegantly incorporates these effects. The compact nature of the associated infinite dimensional K\"ahler manifold secures the correct dimensionality of the the singlet Hilbert space. By using Stirling's approximation for the number of states in the bi-local Hilbert space \eqref{counting}, we see the dimension growing linearly in $N$ (for $K \gg N$)
\begin{eqnarray}
\ln (\dim \mathcal{H}_B) \sim 2 \, N \, K \, \ln 2
\end{eqnarray}
at the leading order. This is a clear demonstration of the presence of an $N$-dependent cutoff in agreement with the fermionic nature of the original $Sp(2N)$ Hilbert space. So in the nonlinear bi-local theory with $G=1/N$ as coupling constant, we have the desired effect that the Hilbert space is cutoff through $1/G$ effects. Consequently we conclude that the geometric bi-local representation with infinite dimensional matrices $Z(k,l)$ provides a complete framework for quantization of the bi-local theory and of de Sitter HS Gravity.

The following further results on quantization of this type of K\"ahler systems have direct relevance to Higher Spin duality. First, the parameter $N$ (and therefore $G$ in Higher Spin Theory) can only take integer values, i.e. $N=0,1,2,3,\cdots$. This feature might appear to be very puzzling from Vasiliev's theory itself, but the fact that there exists a geometric (K\"ahler manifold) representation of the theory provides the explanation. We therefore expect that Vasiliev's theory when suitably canonically quantized takes the form of the above geometric K\"ahler system. We also mention a very recent study of finite $N \rightarrow N+1$ deformation in these theories \cite{Leigh:2012mz}. This can possibly also be investigated by the present Hilbert space method.



\section{Phase Transition}

It was shown by Shenker and Yin \cite{Shenker:2011zf} that the $N$-component vector model undergoes a phase transition at high temperature. The transition occurs at a temperature of order $\sqrt{N}$ and since $1/N=G$ plays the role of coupling constant. This is an important non-perturbative effect characterizes Higher Spin Theories. The argument in \cite{Shenker:2011zf} is based on the exact analysis of $O(N)$ vector model partition function. Here we will show how the transition (and the presence of two phases) can be understood from the bulk field theoretic viewpoint. 

We have already described two versions of bi-local field theory: the covariant one and a canonical (time-like) gauge one. The canonical gauge version (with the Hamiltonian \eqref{H}) represents the singlet spectrum of the theory, and then the  partition function is simply
\begin{eqnarray}
Z(\beta)=\text{Tr}\left( e^{-\beta H^{(2)}} \right)
\end{eqnarray}
giving in the large-$N$ limit the answer
\begin{eqnarray}
F(\beta)=\sum_{\vec{k}_1,\vec{k}_2}\ln \left(1-e^{-\beta \omega(\vec{k}_1,\vec{k}_2)} \right)
\end{eqnarray}
corresponding to the singlet bi-local spectrum with $E_{\vec{k}_1,\vec{k}_2}=\omega(\vec{k}_1,\vec{k}_2)=\vert \vec{k}_1\vert +\vert \vec{k}_2\vert$. This leads to the $O(1)$ result
\begin{eqnarray}
F_1(\beta)\sim V \zeta(5) \, T^4
\end{eqnarray}
where the power (and the argument of the $\zeta$-function) features the dimensionality $D=4$ of the bi-local space: $(\vec{x}_1, \vec{x}_2)$. This recovers the lower phase of \cite{Shenker:2011zf}.

The upper phase can be seen through a stationary point of the bi-local action as it was given in \cite{Das:2003vw}. Namely, the covariant collective action \eqref{action} at finite temperature (with periodic boundary conditions in Euclidean time) has the following stationary-point solution
\begin{eqnarray}
\Phi_\beta(x,y)=\sum_{k_\mu}\frac{e^{i k \cdot (x-y)}}{k_0^2-\vec{k}^2}
\end{eqnarray}
which quantized as $k_0=\frac{2 \pi n}{\beta}$. Evaluation of the action leads to $O(N)$ partition function
\begin{eqnarray}
F_N\equiv S_c(\Phi_\beta)=-\frac{N}{2}\sum_{n,\vec{k}}\ln\left(\vec{k}^2+\left(\frac{2\pi n}{\beta}\right)^2 \right) 
\end{eqnarray}
giving the upper phase result
\begin{eqnarray}
F_N(\beta)\sim -N \, V \zeta(3) \, T^2
\end{eqnarray}
as stated, this result is proportional to $N$ characterizing the $N$-component vector model. In this and also lower phase case the volume is inherited from the spacial volume of the CFT.

An interpretation of this phase transition was suggested by Shenker and Yin in terms of an increase/decrease of number of degrees of freedom, namely from bi-locals to $N$-component partons. From the bi-local field theory viewpoint, we would like to offer an additional interpretation. In terms of the collective dipole (much like in the case of a string) the upper temperature phase is associated with condensation of extra (``winding'') modes, an effect which gives a classical result of order $N=1/G$. The covariant gauge bi-local field (used in exhibiting the upper phase) indeed contains such an extra mode whose relevance comes at finite temperature. It will be interesting to investigate this scenario and in general the physics of this interesting phase transition further.

\ack

The work reviewed in this contribution was done in collaboration with Sumit Das, Diptarka Das, Robert de Mello Koch and Jo\~{a}o Rodrigues. We are grateful to Sumit Das for recent interesting discussions regarding a number of the topics reported. We have also benefited from discussions with Matthias Gaberdiel, Soo-Jong Rey and Mikhail Vasiliev. This work is supported by the Department of Energy under contract DE-FG02-91ER40688.

\end{document}